\documentclass[12pt,english,aps,manuscript]{article}
\usepackage[T1]{fontenc}
\usepackage[latin1]{inputenc}
\usepackage{geometry}
\usepackage{amsmath}
\usepackage[numbers]{natbib}

\makeatletter

\usepackage{geometry}

\makeatletter

\usepackage{color}

\makeatletter
\newcommand{\bee}{\begin{equation}}
\newcommand{\ee}{\end{equation}}
\newcommand{\beea}{\begin{eqnarray}}
\newcommand{\eea}{\end{eqnarray}}

\makeatother

\makeatother

\makeatother

\usepackage{babel}
\begin{document}
\begin{center}
\textbf{\Large Comment on {}``Anomaly Mediation from Unbroken Supergravity'' }
\par\end{center}{\Large \par}

\begin{center}
\vspace{0.3cm}
 
\par\end{center}

\begin{center}
{\large S. P. de Alwis$^{\dagger}$ } 
\par\end{center}

\begin{center}
Physics Department, University of Colorado, \\
 Boulder, CO 80309 USA 
\par\end{center}

\begin{center}
\vspace{0.3cm}
 
\par\end{center}

\begin{center}
\textbf{Abstract} 
\par\end{center}

I argue that ArXiv:1307.3251 is based on a misunderstanding of the
role of Weyl transformations in local off-shell supersymmetry and what
constitutes a true symmetry of the theory. 

\begin{center}
\vspace{0.3cm}
 
\par\end{center}

\vfill{}

$^{\dagger}$dealwiss@colorado.edu

\eject

This note is aimed at clarifying the misconceptions of \citep{D'Eramo:2013mya}.
To begin let us consider some basic facts about the off-shell formulation
of SUGRA. 

The original presentation of this \citep{Cremmer:1982en} in the superspace
version of it given by Wess and Bagger \citep{Wess:1992cp} namely
(with $\kappa=M_{P}^{-1}=1,\, d^{8}z=d^{4}xd^{4}\theta$) is
\begin{eqnarray}
S & = & -3\int d^{8}z{\bf E}\exp[-\frac{1}{3}K(\Phi,\bar{\Phi};Q,\bar{Q}e^{2V})]+\nonumber \\
 &  & \left(\int d^{8}z{\cal E}[W(\Phi,Q)+\frac{1}{4}f(\Phi,Q)){\cal W}{\cal W}]+h.c.\right).\label{eq:action}
\end{eqnarray}
This does not reflect the (super) Weyl invariance of the torsion constraints
of supergravity. In the above the fields $\Phi$, $Q$ are respectively
a set of neutral (for example the moduli of string theory) chiral
superfields and ones charged under the gauge group. $V$ is the gauge
prepotential and ${\bf {\cal W}}{}_{\alpha}=(-\frac{\bar{{\bf \nabla}}^{2}}{4}+2{\bf R})e^{-2V}\nabla_{\alpha}e^{2V}$
is the associated gauge field strength and $\nabla_{\alpha}$ is the
covariant super derivative. Also each term containing non-singlets
is implicitly taken to be an invariant. ${\bf R}$ is the chiral curvature
superfield, ${\bf E}$ is the full superspace measure and ${\cal E}\equiv{\bf E}/2{\bf R}$
is the chiral superspace measure. The torsion constraints of SUGRA
are invariant under Weyl transformations (with a chiral superfield
transformation parameter $\tau$) which are given below.
\begin{eqnarray}
{\bf E}\rightarrow e^{2(\tau+\bar{\tau})}{\bf E}, &  & {\cal E}\,(\equiv\frac{{\bf E}}{2{\bf R}})\rightarrow e^{6\tau}({\cal E}+O(-\frac{{\bf \bar{\nabla}^{2}\bar{\tau}}}{4{\bf R}})),\label{eq:conf1}\\
\nabla_{\alpha}\rightarrow e^{(\tau-2\bar{\tau})}(\nabla_{\alpha}-2(\nabla^{\beta}\tau)M_{\beta\alpha}), &  & V\rightarrow V,\label{eq:conf2}\\
2{\bf R} & \rightarrow & e^{-4\tau}(-\frac{{\bf \bar{\nabla}^{2}}}{4}+2{\bf R)}e^{2\bar{\tau}}\label{eq:conf3}\\
\Phi\rightarrow\Phi,\, Q\rightarrow Q, &  & {\cal W}_{\alpha}\rightarrow e^{-3\tau}{\cal W}_{\alpha}.\label{eq:conf4}
\end{eqnarray}
Here $M_{\alpha\beta}$ is a Lorentz matrix. Clearly \eqref{eq:action}
is not invariant under these transformations.

\[
\]
An action which is manifestly Weyl invariant (see for example \citep{Kaplunovsky:1994fg}
and references therein) is the following,
\begin{eqnarray}
S & = & -3\int d^{8}z{\bf E}C\bar{C}\exp[-\frac{1}{3}K(\Phi,\bar{\Phi};Q,\bar{Q}e^{2V})]+\nonumber \\
 &  & \left(\int d^{8}z{\cal E}[C^{3}W(\Phi,Q)+\frac{1}{4}f(\Phi,Q)){\cal W}{\cal W}]+h.c.\right).\label{eq:action-1}
\end{eqnarray}
This action contains a chiral \textit{scalar} Weyl compensator superfield
$C$ with the transformation rule
\begin{equation}
C\rightarrow e^{-2\tau}C,\label{eq:Ctrans-1}
\end{equation}
and its role is to create a manifestly Weyl invariant action. It can
be hardly overemphasized that $C$ does not contain propagating (i.e.
physical) degrees of freedom and is a redundant superfield. Clearly
it can be gauged away to unity in the classical action. It is merely
a book keeping device that enables one to keep track of the different
Weyl gauges. Quantum mechanically these transformations have an anomaly
since the path integral measure is not invariant. This anomaly was
calculated in \citep{Kaplunovsky:1994fg}(KL). As we review below
this may be dealt with by adding an appropriate correction to the
gauge kinetic term \citep{Kaplunovsky:1994fg}.

From the superspace integration by parts rule and the torsion constraints
\citep{Gates:1983nr,Buchbinder:1998qv} the entire integral can be
written effectively in the same form as the last line of \eqref{eq:action-1},
since 
\[
\int d^{8}z{\bf E}{\bf L}=\int d^{8}z\frac{{\bf E}}{2{\bf R}}(-\frac{1}{4}\bar{{\bf {\bf \nabla}}}{}^{2}+2{\bf R){\bf L},}
\]
where ${\bf L}$ is an arbitrary unconstrained superfield. This relation
enables one to derive superspace equations of motion. Thus varying
the action w.r.t. $C$ gives 
\begin{equation}
-(-\frac{1}{4}\bar{{\bf {\bf \nabla}}}{}^{2}+2{\bf R)}(\bar{C}e^{-K/3})+C^{2}W=0\label{eq:Ceqn}
\end{equation}
In the $C=1$ gauge this equation becomes the set of trace equations
that is obtained by varying with respect to the conformal mode of
the super metric. 

Taking the lowest component of \eqref{eq:Ceqn} gives (with $|_{0}$
an instruction to take the lowest component) 
\begin{equation}
\frac{\bar{F}^{\bar{C}}}{\bar{C}}+2{\bf R}|_{0}=\frac{1}{3}\bar{F}^{\bar{i}}K_{\bar{i}}|_{0}+\frac{C^{2}}{\bar{C}}e^{K/3}W|_{0}+{\rm fermion}\,{\rm terms}\label{eq:FC}
\end{equation}
In this compensator framework the Weyl anomaly needs to be cancelled
and KL do this by making the \emph{replacement} 
\begin{equation}
f_{a}(\Phi,Q)\rightarrow\tilde{f}_{a}(\Phi,Q)\equiv f_{a}(\Phi,Q)-\frac{3c_{a}}{8\pi^{2}}\ln C.\label{eq:f+lnC}
\end{equation}
 Next the field redefinition necessary to get to canonical normalization
for the matter terms needs to be done. We expand to lowest order in
the `MSSM' fields $Q$ and ignore higher than quadratic terms in these
fields since they are expected to get negligible vev's.  This gives
an additional term $-(T_{a}(r)\tau_{Z}/4\pi^{2}$, with%
\footnote{For simplicity we just consider one matter representation $r$.%
}
\begin{equation}
\tau_{Z}+\bar{\tau}_{Z}=\ln(C\bar{C}e^{-\hat{K}/3}\det Z_{r})|_{H}\label{eq:tauZfix}
\end{equation}
Here the instruction on the RHS again requires one to keep just the
harmonic part of the expansion of $K$ (i.e. the sum of the chiral
and anti-chiral parts). Combined with \eqref{eq:f+lnC} this gives
(apart from a term coming from rescaling the gauge kinetic term) the
quantum gauge coupling function at the UV scale%
\footnote{All this is of course essentially contained in \citep{Kaplunovsky:1994fg}.
The explanation of how the anomaly coefficient $c_{a}$ gets replaced
by the beta function coefficient $b_{a}$ was given in \citep{deAlwis:2008aq}. %
},
\begin{eqnarray}
f_{a}^{quantum}+\bar{f}_{a}^{quantum} & \equiv\nonumber \\
f_{a}(\Phi,Q)+\bar{f}_{a}(\bar{\Phi},\bar{Q}) & - & \frac{3c_{a}}{8\pi^{2}}\ln(C\bar{C})-\frac{T_{a}(r)}{4\pi^{2}}\ln(C\bar{C}e^{-K_{m}/3}\det Z_{r})|_{H}\\
 & = & f_{a}(\Phi,Q)+\bar{f}_{a}(\bar{\Phi},\bar{Q})-\frac{b_{a}}{8\pi^{2}}\ln(C\bar{C})-\frac{T_{a}(r)}{4\pi^{2}}\ln(e^{-K_{m}/3}\det Z_{r})|_{H}\label{eq:fquantum}
\end{eqnarray}

The gauge fixing of $C$ to get to the Einstein-Kaehler frame is \citep{Kaplunovsky:1994fg}
\begin{equation}
\ln C+\ln\bar{C}=\frac{K}{3}|_{H}.\label{eq:Cgaugefix}
\end{equation}
This amounts to going to the Wess-Zumino gauge for the real superfield
$K$ and as pointed out by KL is completely equivalent to the set
of transformations done in Wess and Bagger (WB) \citep{Wess:1992cp}
to get to the Einstein-Kaehler frame. In particular in terms of components
\eqref{eq:Cgaugefix} gives,
\begin{eqnarray}
\ln C|_{0}+\ln\bar{C}|_{0} & = & \frac{K}{3}|_{0},\label{eq:scalar}\\
C|_{0}^{-1}\psi_{\alpha}^{(C)} & = & \frac{1}{3}K_{\Phi^{i}}\psi_{\alpha}^{i},\label{eq:fermion}\\
C|_{0}^{-1}F^{C} & = & \frac{1}{3}K_{\Phi^{i}}F^{i}+{\rm fermionic\, terms}.\label{eq:FC-1}
\end{eqnarray}
The objection of \citep{D'Eramo:2013mya} is only to the last equation
above. Actually the authors seem to make two contradictory statements.
Firstly they appear to be claiming that while it is legitimate to
use the first two equations since they are needed to get Einstein
frame action with the properly normalized matter metric, there is
no such necessity to use the F-term equation%
\footnote{Note that using the last equation we have from \eqref{eq:FC} $2{\bf R}|_{0}=e^{K/2}W|_{0}\equiv m_{3/2},$
and this is precisely the value of this auxiliary field that is used
in WB after making a shift to solve for this quantity.%
}. But if one accepts \eqref{eq:Cgaugefix} as a superfield equation
(as one must since the anomaly is a superfield), the second equation
is obtained by applying the operator $\nabla_{\alpha}$ once and taking
the lowest component, and the third is obtained by applying $\nabla^{2}$
and then taking the lowest component. On the other hand the authors
seem perfectly happy to take the F-term of \eqref{eq:tauZfix}, which
has a similar (Wess-Zumino gauge fixing like) character, to get the
Konishi anomaly contribution to the gaugino mass! So it is unclear
what possible objection there could be to this procedure when used
on \eqref{eq:Cgaugefix}.

The main objection however appears to be the claim that while the
above procedure gives the correct result for the gaugino mass term
of the Wilsonian action, this term is {}``gauge dependent''. Let
us take the lowest and the F-components of \eqref{eq:fquantum} but
without fixing the {}``gauge'' \eqref{eq:Cgaugefix}.

\begin{equation}
\frac{1}{g_{a}^{2}(\Phi,\bar{\Phi})}=\Re f_{a}^{quantum}=\Re f_{a}(\Phi)-\frac{b_{a}}{16\pi^{2}}\ln C|_{0}-\frac{T_{a}(r)}{8\pi^{2}}(e^{-K_{m}/3}\ln\det Z_{r})\label{eq:fquantumEgauge}
\end{equation}
and 
\begin{equation}
\frac{2M_{a}}{g_{a}^{2}}(\Phi,\bar{\Phi};\mu)=\frac{1}{2}F^{A}\partial_{A}f_{a}-\frac{b_{a}}{16\pi^{2}}\frac{F^{C}}{C|_{0}}-\sum_{r}\frac{T_{a}(r)}{8\pi^{2}}F^{A}\partial_{A}(-\frac{K_{m}}{3}+{\rm tr}\ln{\bf Z}^{(r)}(g^{2})),\label{eq:MquantumEgauge}
\end{equation}
\citep{D'Eramo:2013mya} argues that the second equation determining
the gaugino mass is gauge dependent since $F^{C}/C$ transforms under
the Weyl transformations \eqref{eq:Ctrans-1}. But this is true of
the first equation as well, since the lowest component of $C$ also
transforms! In other words the authors of \citep{D'Eramo:2013mya}
should have attempted to correct both the gaugino mass and the expression
for the gauge coupling in KL, since both are physical quantities and
should be {}``gauge independent''. What these authors want to do
is to replace the RHS of \eqref{eq:FC-1} by the RHS of \eqref{eq:Ceqn}
by arbitrarily choosing ${\bf R}|_{0}=0$. But this choice violates
off-shell supersymmetry since there is no way in which the supertransformations
of the RHS of \eqref{eq:scalar}\eqref{eq:fermion} (which are defined
in terms of the chiral scalar supermultiplets) can acquire a term
which is proportional to the F-term of an independent multiplet -
namely the gravity multiplet. 

The argument in \citep{D'Eramo:2013mya} is based on a misunderstanding
of the role of Weyl transformations. A true symmetry acts only on
the physical propagating fields and under such symmetry transformations
the action is invariant. But \eqref{eq:action-1} has been made invariant
under Weyl transformations only after the introduction of an additional
auxiliary non-propagating field $C$. There is absolutely no reason
to demand that physical quantities are invariant under Weyl transformations.
The introduction of $C$ is merely a convenient way of representing
the effect of the change in the Weyl frame - in our case that of going
from the original Jordan frame to the final Einstein frame. These
transformations produce effects coming from the measure. The introduction
of the $\ln C$ term is merely one way of representing them. 

Alternatively as pointed out by KL and discussed at considerable length
in the Appendix of \citep{deAlwis:2012gr}, the results can be derived
without ever introducing the auxiliary field $C$! Here one simply
picks up the extra terms from the anomaly upon doing the appropriate
transformations to get to the Einstein-Kaehler gauge. Indeed this
is how the Konishi terms are obtained anyway in both methods! 

If the authors are right then every anomaly can be got rid of in this
way by adding an additional non-propagating field. Thus $\pi^{0}$
should not decay in the chiral limit and there would be no running
of physical couplings in (massless) QCD etc! Anomalies have real physical
effects and it is precisely because the action is not invariant under
an anomalous symmetry that these effects are manifested. This has
nothing to do with the difference between Wilsonian and 1PI effective
actions. In fact the 1PI action corresponding to \eqref{eq:action}
should display the full effect of the anomaly (since it includes the
integration over the massless fields). In other words its super-Weyl
variation should simply give the super-Weyl anomaly including its
F-term. So the additional contribution to the gauge coupling and the
gaugino mass (after doing also the additional field redefinitions
needed to get canonical normalization for the charged matter fields),
should be precisely what is given by \eqref{eq:fquantumEgauge} and
\eqref{eq:MquantumEgauge}.

In summary the anomalous contribution to both the gauge coupling and
the associated gaugino mass can be obtained without introducing the
auxiliary super field $C$. In this case it is obvious that Weyl invariance
is a red-herring and that the real issue is the anomaly coming from
changing the frame from the original SUGRA one to the Einstein-Kähler
frame. The KL formulae can be obtained without any confusion
about the role compensators. The introduction of $C$ is merely a
device for discussing the change of frame and this takes the form
two different choices for it; $C=1$ for the original SUGRA (or Jordan)
frame given in WB for instance, and \eqref{eq:Cgaugefix} for the
Einstein-Kähler frame. These two are not equivalent because of the
anomaly and there is an anomalous contribution to the coupling as
well as the gaugino mass. One cannot arbitrarily change the F-term
as done in \citep{D'Eramo:2013mya} without violating supersymmetry
and/or general covariance. It makes no sense to impose the Weyl symmetry
on the gauge fixed object and to treat the lowest component (the coupling
constant) and the F-term (gaugino mass) differently.

I wish to acknowledge discussions with Jan Louis. This research is
partially supported by the United States Department of Energy under
grant DE-FG02-91-ER-40672. 

\[
\]
 \bibliographystyle{apsrev} \bibliographystyle{apsrev}
\bibliography{myrefs}

\end{document}